\documentclass[10pt]{revtex4}
\usepackage{amssymb}
\usepackage{latexsym}
\usepackage{epsfig}
\begin{document}
 
\title{The BIonic diode in a system of trigonal manifolds}

\author{Alireza Sepehri $^{1}$\footnote{alireza.sepehri@uk.ac.ir}}

 \affiliation{ $^{1}$ Research Institute for Astronomy and Astrophysics of
Maragha (RIAAM), P.O. Box 55134-441, Maragha, Iran.}

\author{Mohd.Zeyauddin$^{2}$\footnote{mdzeyauddin@gmail.com}}

\affiliation{$^{2}$Department of General Studies (Mathematics Section), Jubail Industrial College, 
Royal Commission for Jubail and Yanbu, Jubail Industrial City 31961, Kingdom of Saudi Arabia}

\author{Anirudh Pradhan$^{3}$\footnote{pradhan.anirudh@gmail.com}}

\affiliation{$^{3}$Department of Mathematics, Institute of Applied Sciences and Humanities, G L A
 University, Mathura-281 406, Uttar Pradesh, India.}

\begin{abstract}
A BIonic diode is constructed of two polygonal manifolds connected by a  Chern-Simons manifold. 
The shape and the angle between atoms of molecules on the boundary of two polygonal manifolds  are completely 
different. For this reason, electrons on the Chern-Simons manifold are repelled by  molecules at the boundary 
of  one manifold and absorbed by  molecules on the boundary of another manifold. The attractive and repulsive 
forces between electrons are carried by masive photons.  For example, when two non-similar trigonal manifolds 
join to each other, one non-symmetrical hexagonal manifold is emerged and the exchanged photons form Chern-Simons 
fields which live on a Chern-Simons  manifold in a BIon. While, for a hexagonal manifold, with similar trigonal 
manifolds, the photons exchanged between two trigonal manifolds cancel the effect of each other and BIonic energy 
becomes zero. Also, exchanging photons between heptagonal and pentagonal manifolds lead to the motion of electrons 
on the Chern-Simons manifold and formation of BIonic diode.  The mass of  photons depend on the shape of molecules 
on the boundary of manifolds and the length of BIon in a gap between two manifolds. 
\vspace{5mm}\noindent\vspace{0.8mm}\newline Keywords: Diode, Photons, Holography, Chern-Simons, BIonic system
\end{abstract}

\maketitle
\section{Introduction}

A diod is constructed of one subsystem with extra electrons which are paired with extra holes in 
other subsystem. By applying one external force, these pairs are broken and an electrical current is
produced. Until now, less discussions have been done on this subject. For example - the researches of 
the past few years have shown that graphene can build junctions with $3D$ or $2D$ semi-conductor
materials which have rectifying characteristics and act as excellent Schottky diodes \cite{q2}. The 
main novelty of these systems is the tunable Schottky barrier height-a property which makes the 
graphene/semiconductor junction a great platform for the consideration of interface transport methods, as 
well as using in photo-detection \cite{q3}, high-speed communications \cite{q4}, solar cells \cite{q5}, 
chemical and biological sensing \cite{q6}, etc. Also, discovering an optimal Schottky interface of graphene, 
on other matters like $Si$, is challenging, as the electrical transport is corresponded on the graphene 
quality and the temperature.  Such interfaces are of increasing research hope for integration in diverse electronic 
systems being thermally and chemically stable in all environments, unlike standard metal/semiconductor 
interfaces \cite{q7}. \\

Previously, we have considered the process of formation of a holographic diode by joining polygonal manifolds \cite{q8}. 
In our model, first a big manifold with polygonal molecules is broken, two child manifolds and one Chern-Simons manifold 
appeared. Then, heptagonal molecules on one of child manifolds repel electrons  and pentagonal molecules on another 
child manifold absorb them. Since, the angle between atoms in heptagonal molecules with respect to center of 
that is less than  pentagonal molecules and  parallel electrons come nearer to each other and in terms of Pauli exclusion 
principle, therefore are repelled. Also, parallel electrons in pentagonal molecules become more distant and some holes emerge. 
Consequently, electrons move from one of child manifolds with heptagonal molecules towards other child manifold with 
pentagonal molecules via the Chern-Simons manifold and a diode emerges. Also, we have discussed that this is a 
real diod that may be built by bringing heptagonal and pentagonal molecules among the hexagonal molecules of graphene. 
To construct this diode, two graphene sheets are needed which are connected through a tube. Molecules, at the 
junction points of one side of the tube, should have the heptagonal shapes and other molecules at the junction 
points of another side of the tube should have the pentagonal shapes. Heptagonal molecules repel and pentagonal 
molecules absorb electrons and a current between two sheets is produced. This current is very similar to the current 
which is produced between layers $N$ and $P$ in a system in solid state. This current was produced only from the side 
with heptagonal molecules towards the side with pentagonal molecules. Also, the current from the sheet with pentagonal 
molecules towards the sheet with heptagonal molecules is zero. This characteristic can also be seen in normal diod. \\

In this paper, we extend the consideration on holographic diodes to BIonic systems. A BIon is a system which consist 
of two polygonal manifolds connected by a Chern-Simons manifold. We will show that when two manifolds with 
two different types of polygonal molecules come close to each other, some massive photons appear. These photons 
join to each other and build Cherns-Simons fields. These fields lead to the motion of electrons on the Chern-Simons 
manifold between two manifolds and one BIonic diode emerges. The mass of these photons depend on the shape of molecules 
on the manifolds and the length of gap. From this point of view, our result is consistent with previous predictions for 
the mass of photons in \cite{qq8}. \\

The outline of the paper is as follows: In section \ref{o1}, we will show that by joining non-similar trigonal manifolds, a 
hexagonal diode emerges. In this diode, photons join to each other and form the Chern-Simons  fields. In section \ref{o2}, 
we will consider the process of the formation of a BIonic diode from a manifold with heptagonal molecules, a manifold with 
pentagonal molecules and a Chern-Simons manifold. We will show that exchanged photons between manifold are massive and their 
mass depends on the length of gap between manifolds. The last section is devoted to summary and conclusion.

\section{The hexagonal diode }\label{o1}

In this section, we will show that a hexagonal diode can be built by joining two non-similar trigonal manifolds which exchanges 
photons between them form Chern-Simons fields. These fields force electrons and lead to their motion between two trigonal 
manifolds. Also, we will explain that if two similar trigonal manifolds join to each other, exchanged photons cancel the 
effect and no diode emerges. \\

Previously, in ref \cite{q8}, for explaining graphene systems, we have used of the concept of string theory. In our model, 
scalar strings ($X$) are produced by pairing two electrons with up and down spins. Also, $A$ denotes the photon which is 
exchanged between electrons and $F$ is the photonic field strength. Now, we will extend this model to polygonal manifolds and 
trigonal manifolds and write the following action   \cite{q8,D3,Df}:

\begin{eqnarray}
&& S_{3}=-T_{tri} \int d^{3}\sigma \sqrt{\eta^{ab}
	g_{MN}\partial_{a}X^{M}\partial_{b}X^{N}+2\pi
	l_{s}^{2}G(F))}\nonumber\\&&
G=(\sum_{n=1}^{3}\frac{1}{n!}(-\frac{F_{1}..F_{n}}{\beta^{2}}))
\nonumber\\&& F=F_{\mu\nu}F^{\mu\nu}\quad
F_{\mu\nu}=\partial_{\mu}A_{\nu}-
\partial_{\nu}A_{\mu}\label{f1}
\end{eqnarray}

where $g_{MN}$ is the background metric, $ X^{M}(\sigma^{a})$'s
are scalar fields which are produced by pairing electrons, $\sigma^{a}$'s are
the manifold coordinates, $a, b = 0, 1, ..., 3$ are world-volume
indices of the manifold and $M,N=0, 1, ..., 10$ are eleven dimensional
spacetime indices. Also, $G$ is the nonlinear field \cite{Df} and $A$ is the photon 
which exchanges between electrons. Using the above action, the  Lagrangian for trigonal 
manifold can be written as:
\begin{eqnarray}
&&\L=-4\pi T_{tri} \int d^{3}\sigma \sqrt{1+(2\pi
	l_{s}^{2})^{2}G(F)+ \eta^{ab}g_{MN}\partial_{a}X^{M}\partial_{b}X^{N}}\label{f2} \;,
\end{eqnarray}

where prime denotes the derivative respect to $\sigma$. To derive
the Hamiltonian, we have to obtain the canonical momentum density
for photon. Since we are interesting to consider electrical
solutions. Therefore we suppose that $F_{01}\neq 0$ and other components of
$F_{\alpha \beta}$ are zero. So, we have
\begin{eqnarray}
&&\Pi=\frac{\delta \L}{\delta
	\partial_{t}A^{1}}=-\frac{\sum_{n=1}^{3}\frac{n}{n!}(-\frac{F_{1}..F_{n-1}}{\beta^{2}})F_{01}}{\beta^{2}\sqrt{1+(2\pi
		l_{s}^{2})^{2}G(F)+ \eta^{ab}g_{MN}\partial_{a}X^{M}\partial_{b}X^{N}}} \label{f3}
\end{eqnarray}
so, by replacing $\int d^{3}\sigma =\int d\sigma \sigma^{2}$, the Hamiltonian may be built as \cite{D3,Df}:
\begin{eqnarray}
&&H=4\pi T_{tri}\int d\sigma
\sigma^{2}\Pi\partial_{t}A^{1}-\L=4\pi\int d\sigma [
\sigma^{2}\Pi(F_{01})-\partial_{\sigma}(\sigma^{2}\Pi)A_{0}]-\L
\label{f4}
\end{eqnarray}

In the second step, we use integration by parts and obtain the
term proportional to $\partial_{\sigma}A$. Using the constraint
($\partial_{\sigma}(\sigma^{2}\Pi)=0$), we obtain \cite{D3}:
\begin{eqnarray}
&& \Pi=\frac{k}{4\pi \sigma^{2}}, \label{f5}
\end{eqnarray}
where $k$ is a constant. Substituting equation (\ref{f5}) in
equation (\ref{f4}) and $\int d^{3}\sigma =\int d\sigma_{3} d\sigma_{2} d\sigma_{1} $  yields the following Hamiltonian:
\begin{eqnarray}
&&H_{1}=4\pi T_{tri}\int d\sigma_{3} d\sigma_{2} d\sigma_{1} \sqrt{1+(2\pi
l_{s}^{2})^{2}\sum_{n}\frac{n}{n!}(-\frac{F_{1}..F_{n-1}}{\beta^{2}})+\eta^{ab}g_{MN}\partial_{a}X^{M}\partial_{b}X^{N}}O_{1}
\nonumber\\&&O_{1}=\sqrt{1+\frac{k^{2}_{1}}{\sigma^{4}_{1}}}
\label{f6}
\end{eqnarray}

To obtain the explicit form of wormhole- like tunnel which goes out of trigonal manifold, we need 
a Hamiltonian in terms of separation distance
between sheets. For this reason, we redefine Lagrangian as:
\begin{eqnarray}
&&\L=-4\pi T_{tri} \int d\sigma \sigma^{2}\sqrt{1+(2\pi
	l_{s}^{2})^{2}\sum_{n}\frac{n}{n!}(-\frac{F_{1}..F_{n-1}}{\beta^{2}})+z'^{2}}O_{1}\label{f7}
\end{eqnarray}
With this new form of Lagrangian, we repeat our previous
calculations. We have
\begin{eqnarray}
&&\Pi=\frac{\delta \L}{\delta
	\partial_{t}A^{1}}=-\frac{\sum_{n}\frac{n(n-1)}{n!}(-\frac{F_{1}..F_{n-2}}{\beta^{2}})F_{01}F_{1}}{\beta^{2}\sqrt{1+(2\pi
		l_{s}^{2})^{2}\sum_{n}\frac{n}{n!}
		(-\frac{F_{1}..F_{n-1}}{\beta^{2}})+\eta^{ab}g_{MN}\partial_{a}X^{M}\partial_{b}X^{N}}}
\label{f8}
\end{eqnarray}

So the new Hamiltonian can be constructed as:
\begin{eqnarray}
&&H_{2}=4\pi T_{tri}\int d\sigma
\sigma^{2}\Pi\partial_{t}A^{1}-\L=4\pi\int d\sigma [
\sigma^{2}\Pi(F_{01})-\partial_{\sigma}(\sigma^{2}\Pi)A_{0}]-\L
\label{f9}
\end{eqnarray}

Again, we use integration by parts to obtain the term proportional
to $\partial_{\sigma}A$. Imposing the constraint
($\partial_{\sigma}(\sigma^{2}\Pi)=0$), we obtain:
\begin{eqnarray}
&& \Pi=\frac{k}{4\pi \sigma^{2}} \label{f10}
\end{eqnarray}
where k is a constant. Substituting equation (\ref{f10}) in
equation (\ref{f9}) yields the following Hamiltonian:
\begin{eqnarray}
&&H_{2}=4\pi T_{tri}\int d\sigma_{3} d\sigma_{2} d\sigma_{1}  \sqrt{1+(2\pi
	l_{s}^{2})^{2}\sum_{n}\frac{n(n-1)}{n!}(-\frac{F_{1}..F_{n-2}}{\beta^{2}})+\eta^{ab}g_{MN}\partial_{a}X^{M}\partial_{b}X^{N}}
O_{2}\nonumber\\
&&O_{2}=O_{1}
\sqrt{1+\frac{k^{2}_{2}}
	{O_{1}\sigma^{4}_{2}}} \label{f11}
\end{eqnarray}

And if we repeat these calculations for 3 times, we obtain
\begin{eqnarray}
&&H_{3}=4\pi T_{tri}\int d\sigma_{3} d\sigma_{2} d\sigma_{1} \sqrt{1+\eta^{ab}g_{MN}\partial_{a}X^{M}\partial_{b}X^{N}}O_{tot}
\nonumber\\ &&
O_{tot}=\sqrt{1+\frac{k^{2}_{3}}{O_{2}\sigma^{4}_{3}}}\sqrt{1+\frac{k^{2}_{2}}{O_{1}\sigma^{4}_{2}}}\sqrt{1+\frac{k^{2}_{1}}{\sigma^{4}_{1}}}
\nonumber\\
&&O_{2}=O_{1}\sqrt{1+\frac{k^{2}_{2}}{O_{1}\sigma^{4}_{2}}}
\label{f12}
\end{eqnarray}

At this stage, we will make use of some  approximations and obtain the simplest form of the  Hamiltonian  of trigonal manifold: 

\begin{eqnarray}
&&A\sqrt{1+\frac{k^{2}}{O_{1}\sigma^{4}}}\sqrt{1+\frac{k^{2}}{\sigma^{4}}}\simeq
A\sqrt{1+\frac{k^{2}}{\sigma^{4}}}+A\frac{k^{2}}{2\sigma^{4}}\simeq
\nonumber\\ &&
A+A\frac{k^{2}}{2\sigma^{4}}+A\frac{k^{2}}{2\sigma^{4}}=\frac{A}{2}(1+\frac{k^{2}}{\sigma^{4}})+
\frac{A}{2}(1+\frac{k^{2}}{\sigma^{4}})\simeq \nonumber\\ &&
2A'\sqrt{1+\frac{k^{2}}{\sigma^{4}}}\Rightarrow
O_{tot}=\frac{3}{2}\sqrt{1+\frac{k^{2}}{\sigma^{4}}}=\frac{3}{2}
O_{1}\Rightarrow \nonumber\\ &&
 H_{3}=4\pi T_{tri}\int d\sigma
\sigma^{2} \sqrt{1+\eta^{ab}g_{MN}\partial_{a}X^{M}\partial_{b}X^{N}}O_{tot}=4\pi 3 T_{tri}\int d\sigma
\sigma^{2} \sqrt{1+\eta^{ab}g_{MN}\partial_{a}X^{M}\partial_{b}X^{N}}O_{1}=\nonumber\\ &&4\pi 3 T_{tri}\int
d\sigma \sigma^{2}
\sqrt{1+\eta^{ab}g_{MN}\partial_{a}X^{M}\partial_{b}X^{N}}\sqrt{1+\frac{k^{2}}{\sigma^{4}}}=\frac{3}{2}H_{linear}\nonumber\\ &&\nonumber\\
&&H_{linear}=4\pi 3 T_{tri}\int d\sigma \sigma^{2}
\sqrt{1+\eta^{ab}g_{MN}\partial_{a}X^{M}\partial_{b}X^{N}}\sqrt{1+\frac{k^{2}}{\sigma^{4}}}\label{f16}
\end{eqnarray}
where $A'=\frac{A}{2}$ is a constant that depends on the trigonal manifold
action($T_{tri}$)and other stringy constants. This equation shows
that each pair of electrons on each side of trigonal manifold connected by a wormhole- like tunnel and form a linear
BIon; then these BIons join to each other and construct a 
nonlinear  trigonal BIon. 

For constrcuting a hexagonal manifold, we should put two trigonal manifolds near each other so that direction of the 
motion of electrons and photons on two trigonal manifolds are reverse to each other ( Fig.1.). In a symmetrical 
hexagonal manifold, two photons cancel the effect of each other and total energy of system becomes zero. Using expressions given in
Eq. (\ref{f12}), we can write:
   
\begin{eqnarray}
&& 
\sigma_{1}\rightarrow -\bar{\sigma}_{1} \quad \sigma_{2}\rightarrow -\bar{\sigma}_{2} \quad \sigma_{3}\rightarrow -\bar{\sigma}_{3} 
\nonumber\\&& \int d\sigma_{3} d\sigma_{2} d\sigma_{1}\rightarrow   
-\int d\bar{\sigma}_{3} d\bar{\sigma}_{2} d\bar{\sigma}_{1}\nonumber\\&& A_{0} \rightarrow \bar{A}_{0} \quad A_{1} 
\rightarrow \bar{A}_{1}\nonumber\\&&\Rightarrow H_{3}\rightarrow -\bar{H}_{3} \label{EQ1}
\end{eqnarray}

For a symmetrical hexagonal manifold, the Hamiltonians of two trigonal manifolds cancel the effect of each other and total 
Hamiltonian of system becomes zero. This system is completely stable and can't interact with other systems. For a non-symmetrical 
hexagonal manifold, fields are completely different and two Hamiltonian cannot cancel the effect of each other. Using equations 
(\ref{f1} and \ref{f12} ), we have: 

\begin{eqnarray}
&&H_{3}=4\pi T_{tri}\int d\sigma_{3} d\sigma_{2} d\sigma_{1} \sqrt{1+\eta^{ab}g_{MN}\partial_{a}X^{M}\partial_{b}X^{N}}O_{tot}
\nonumber\\ &&
\neq \bar{H}_{3}=4\pi T_{tri}\int d\bar{\sigma}_{3} d\bar{\sigma}_{2} d\bar{\sigma}_{1} \sqrt{1+\eta^{ab}g_{MN}
\partial_{a}\bar{X}^{M}\partial_{b}\bar{X}^{N}}\bar{O}_{tot} \nonumber\\ && \Rightarrow S=-T_{tri} \int d^{3}\sigma \sqrt{\eta^{ab}
	g_{MN}\partial_{a}X^{M}\partial_{b}X^{N}+2\pi
	l_{s}^{2}G(F))}\nonumber\\ &&
\neq  \bar{S}=-T_{tri} \int d^{3}\bar{\sigma} \sqrt{\eta^{ab}
	g_{MN}\partial_{a}\bar{X}^{M}\partial_{b}\bar{X}^{N}+2\pi
	l_{s}^{2}G(\bar{F}))}
\label{EQ2}
\end{eqnarray}

Thus, total Hamiltonian and the action of two trigonal manifolds can be obtained as:

\begin{eqnarray}
	&&H_{6}^{tot}=H_{3}-\bar{H}_{3}\nonumber\\&& S_{6}^{tot}=S_{3}-\bar{S}_{3}
	\label{EQ3}
\end{eqnarray}

This equation shows that if two trigonal manifolds join to each other and form the hexagonal manifold, the Hamiltonian and 
also the action of hexagonal manifold is equal to the difference between the actions and Hamiltonians of two trigonal manifolds. 
A non-symmetrical hexagonal manifold has an active potential and can interact with other manifolds (See Fig.2.). \\

At this stage, we can assert that the exchanged photons between two trigonal manifolds produce the Chern-Simons fields. 
To write our model in terms of concepts in supergravity, we should define G and C-fields. G- fields with four indices 
are constructed from two strings and C-fields with three indices are produced when three ends of G-fields are placed 
on one manifold and one index is located on one another manifold (Figure 3). We can define G and Cs-fields as follows:

\begin{eqnarray}
&& G_{IJKL}\approx F_{[IJ}F_{KL]} \nonumber\\&& Cs_{IJK}=
\epsilon^{IJK} F_{IJ}A_{K}
\label{EQ4}
\end{eqnarray} 
To obtain G- and  Cs-fields, we will assume that two spinors with up and down spins couple to each other and exchanged 
photons ($X^{M}\rightarrow A^{M}\psi_{\downarrow}\psi_{\uparrow}$, $X^{0}\rightarrow t$). We also assume that spinors 
are only functions  of coordinates ($\sigma$, $t$). Using equation (\ref{f3}, \ref{f5},\ref{f12}), we obtain:                                                                                                                                                                                               
\begin{eqnarray}
&& \Pi=\frac{k}{4\pi \sigma^{2}}\nonumber\\&&
=\frac{\sum_{n=1}^{3}\frac{n}{n!}
	(-\frac{\bar{F}_{1}..\bar{F}_{n-1}}
	{\beta^{2}})\bar{F}_{01}}{\beta^{2}\sqrt{1 + ([\bar{A}^{M}\bar{A}_{M}(\psi_{1,\downarrow}
	\psi_{1,\uparrow}\psi_{2,\downarrow}\psi_2,{\uparrow})]')^{2}+2\pi
			l_{s}^{2}(\sum_{n=1}^{3}\frac{1}{n!}(-\frac{\bar{F}_{1}..\bar{F}_{n}}{\beta^{2}}))}}  \nonumber\\&& 
			H_{3}=4\pi T_{tri}\int d\sigma_{3} d\sigma_{2} d\sigma_{1} \sqrt{1 + ([A^{M}A_{M}(\psi_{1,\downarrow}
			\psi_{1,\uparrow}\psi_{2,\downarrow}\psi_2,{\uparrow})]')^{2}
	}O_{tot}\nonumber\\&& H_{3}=4\pi T_{tri}\int d\sigma_{3} d\sigma_{2} d\sigma_{1} \sqrt{1 + 
	([A^{M}A_{M}(\psi_{1,\downarrow}\psi_{1,\uparrow}\psi_{2,\downarrow}\psi_2,{\uparrow})]')^{2}}\times\nonumber\\ &&
\sqrt{1+\frac{1}{O_{2}}(\frac{\sum_{n=1}^{3}\frac{n}{n!}
		(-\frac{\bar{F}_{1}..\bar{F}_{n-1}}
		{\beta^{2}})\bar{F}^{3}_{01}}{\beta^{2}\sqrt{1 + ([\bar{A}^{M}\bar{A}_{M}(\psi_{1,\downarrow}
	\psi_{1,\uparrow}\psi_{2,\downarrow}\psi_2,{\uparrow})]')^{2}+2\pi
			l_{s}^{2}(\sum_{n=1}^{3}\frac{1}{n!}(-\frac{\bar{F}_{1}..\bar{F}_{n}}{\beta^{2}}))}})^{2}}\times\nonumber\\&&
			\sqrt{1+\frac{1}{O_{1}}(\frac{\sum_{n=1}^{3}\frac{n}{n!}
		(-\frac{\bar{F}_{1}..\bar{F}_{n-1}}
	{\beta^{2}})\bar{F}_{01}^{2}}{\beta^{2}\sqrt{1 + ([\bar{A}^{M}\bar{A}_{M}(\psi_{1,\downarrow}\psi_{1,\uparrow}\psi_{2,\downarrow}
	\psi_2,{\uparrow})]')^{2}+2\pi
	l_{s}^{2}(\sum_{n=1}^{3}\frac{1}{n!}(-\frac{\bar{F}_{1}..\bar{F}_{n}}{\beta^{2}}))}})^{2}}\times\nonumber\\&&
	\sqrt{1+(\frac{\sum_{n=1}^{3}\frac{n}{n!}
(-\frac{\bar{F}_{1}..\bar{F}_{n-1}}
{\beta^{2}})\bar{F}_{01}^{1}}{\beta^{2}\sqrt{1 + ([\bar{A}^{M}\bar{A}_{M}(\psi_{1,\downarrow}\psi_{1,\uparrow}\psi_{2,\downarrow}
\psi_2,{\uparrow})]')^{2}+2\pi
l_{s}^{2}(\sum_{n=1}^{3}\frac{1}{n!}(-\frac{\bar{F}_{1}..\bar{F}_{n}}{\beta^{2}}))}})^{2}}\label{EQ5}
\end{eqnarray}

Where we have shown the exchanged photons on trigonal manifold by ($A,F$) and the exchanged photons on a gap between two 
trigonal manifolds by ($\bar{A},\bar{F}$). By using the Taylor expansion method and substituting results of  (\ref{EQ4}) 
in equation (\ref{EQ5}), we obtain:                                                                                                                                

\begin{eqnarray}
&& H_{tot}^{6} = H_{3} - \bar{H}_{3} \approx
 \nonumber\\&& (4\pi T_{tri})[1 + (\frac{2\pi
	l_{s}^{2}}{\beta^{2}})\bar{F}_{[IJ}\bar{F}_{KL]}\bar{F}^{[IJ}F^{KL]} +(\frac{(2\pi
	l_{s}^{2})^{2}}{\beta^{4}})(\psi_{1,\downarrow}\psi_{1,\uparrow}\psi_{2,\downarrow}\psi_2,{\uparrow})' 
	\epsilon^{IJK} \bar{F}_{IJ}A_{K}\bar{F}_{[IJ}\bar{F}_{KL]}\bar{F}^{[IJ}\bar{F}^{KL]}   \nonumber\\&& -(\frac{2\pi
	l_{s}^{2}}{\beta^{2}})^{3}
\bar{F}_{[IJ}\bar{F}_{KL}\bar{F}_{MN]}\bar{F}^{[IJ}F^{KL}\bar{F}^{MN]} -  (\frac{(2\pi
	l_{s}^{2})^{2}}{\beta^{4}})^{3}(\psi_{1,\downarrow}\psi_{1,\uparrow}\psi_{2,\downarrow}\psi_2,{\uparrow})' \epsilon^{IJK} \bar{F}_{IJ}A_{K}
\bar{F}_{[IJ}\bar{F}_{KL}\bar{F}_{MN]}\bar{F}^{[IJ}F^{KL}\bar{F}^{MN]} +....]\nonumber\\&&- (4\pi T_{tri})[1 + (\frac{2\pi
	l_{s}^{2}}{\beta^{2}})\bar{F}_{[IJ}\bar{F}_{KL]}\bar{F}^{[IJ}F^{KL]} +(\frac{(2\pi
	l_{s}^{2})^{2}}{\beta^{4}})(\psi_{1,\downarrow}\psi_{1,\uparrow}\psi_{2,\downarrow}\psi_2,{\uparrow})' 
	\epsilon^{IJK} \bar{F}_{IJ}A_{K}\bar{F}_{[IJ}\bar{F}_{KL]}\bar{F}^{[IJ}\bar{F}^{KL]}   \nonumber\\&& -(\frac{2\pi
	l_{s}^{2}}{\beta^{2}})^{3}
F_{[IJ}F_{KL}F_{MN]}F^{[IJ}F^{KL}F^{MN]} -  (\frac{(2\pi
	l_{s}^{2})^{2}}{\beta^{4}})^{3}(\psi_{1,\downarrow}\psi_{1,\uparrow}\psi_{2,\downarrow}\psi_2,{\uparrow})' 
	\epsilon^{IJK} F_{IJ}\bar{A}_{K}
F_{[IJ}F_{KL}F_{MN]}F^{[IJ}F^{KL}F^{MN]} +....]\nonumber\\&& =(4\pi T_{tri}) [1+(\frac{2\pi
	l_{s}^{2}}{\beta^{2}}) [\bar{G}_{IJKL}\bar{G}^{IJKL}-G_{IJKL}G^{IJKL}]\nonumber\\&& +
(\frac{(2\pi 
	l_{s}^{2})^{2}}{\beta^{4}})(\psi_{1,\downarrow}\psi_{1,\uparrow}\psi_{2,\downarrow}\psi_2,{\uparrow})'[ Cs \bar{G}_{IJKL}
\bar{G}^{IJKL}-\bar{Cs} G_{IJKL}
G^{IJKL} ]\nonumber\\&&
\nonumber\\&& -(\frac{2\pi
	l_{s}^{2}}{\beta^{2}})^{3} [\bar{G}_{IJKLMN}\bar{G}^{IJKLMN}-G_{IJKLMN}G^{IJKLMN}]\nonumber\\&& -
(\frac{(2\pi 
	l_{s}^{2})^{2}}{\beta^{4}})^{3}(\psi_{1,\downarrow}\psi_{1,\uparrow}\psi_{2,\downarrow}\psi_2,{\uparrow})'[ Cs \bar{G}_{IJKLMN}
\bar{G}^{IJKLMN}-\bar{Cs} G_{IJKLMN}
G^{IJKLMN} ]+.........]
\label{EQ6}
\end{eqnarray}

This equation shows that exchanged photons join to each other and build Chern-Simons fields. These fields make a bridge between two 
trigonal manifolds and produce the BIonic diode  (Figure 4.). For two similar trigonal manifolds, total Hamiltonian of BIon is 
zero, while for two different trigonal manifolds, a BIon is emerged. This BIon is a bridge for transferring energy of one manifold 
to the other. 
 
 At this stage, we can obtain the mass of exchanged photons between two trigonal manifolds. The length of photon relates to the 
 separation distance between electrons or the length of Chern-Simons manifold and the mass of photon depends on the coupling 
 between electrons ($m^{2}=(\psi_{1,\downarrow}\psi_{1,\uparrow}\psi_{2,\downarrow}\psi_2,{\uparrow})$). The equation of motion 
 for $[A^{M}A_{M}(\psi_{1,\downarrow}\psi_{1,\uparrow}\psi_{2,\downarrow}\psi_2,{\uparrow})]'$ which is extracted from the 
 Hamiltonian of (\ref{EQ5}) is
 \begin{eqnarray}
 && A^{M}A_{M}(\psi_{1,\downarrow}\psi_{1,\uparrow}\psi_{2,\downarrow}\psi_2,{\uparrow})\rightarrow m_{photon}^{2}l_{Chern-Simons}
 \nonumber\\&&[m_{photon}^{2}l_{Chern-Simons}]'= (\frac{[O_{tot}(\sigma)^{2}-\bar{O}_{tot}(\sigma)^{2}]}{[O_{tot}
 (\sigma_{0})^{2}-\bar{O}_{tot}(\sigma_{0})^{2}]}-1)^{-1/2}
 \label{EQ7}
 \end{eqnarray}
 
 Solving this equation, we obtain:
 \begin{eqnarray}
 &&[m_{photon}^{2}]=\frac{1}{l_{Chern-Simons}} \int_{\sigma}^{\infty}
 d\sigma'(\frac{[O_{tot}(\sigma)^{2}-\bar{O}_{tot}(\sigma)^{2}]}{[O_{tot}(\sigma_{0})^{2}-\bar{O}_{tot}(\sigma_{0})^{2}]}-1)^{-1/2}
 \label{EQ8}
 \end{eqnarray}
 
Eq. (\ref{EQ8}) shows that photonic mass depends on the length of Chern-Simons manifold and also the length of trigonal manifolds. 
This result is in agreement with previous predictions in \cite{qq8} that photonic mass depends on the parametters of a gap between two systems. 

%%%%%%%%%%%%%%%%%%%%%%%%%%%%%%%%%%%%%%%%%%%%%%%%%%% Figure 1  %%%%%%%%%%%%%%%%%%%%%%%%%%%%
\begin{figure*}[thbp]
	\begin{center}
		\begin{tabular}{rl}
			\includegraphics[width=5cm]{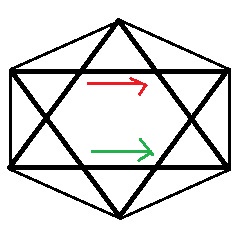}
		\end{tabular}
	\end{center}
	\caption{A symmetric hexagonal manifold is formed by joining two similar trigonal manifolds.The direction of photons 
	on two trigonal manifolds are reverse to each other and they cancel the effect of each other. }
\end{figure*}
%%%%%%%%%%%%%%%%%%%%%%%%%%%%%%%%%%%%%%%%%%%%%%%%%%%%%%%%%%%%%%%%%%%%%%%%%%%%%%%%%%%%%%% 

%%%%%%%%%%%%%%%%%%%%%%%%%%%%%%%%%%%%%%%%%%%%%%%%%%% Figure 2  %%%%%%%%%%%%%%%%%%%%%%%%%%%%
\begin{figure*}[thbp]
	\begin{center}
		\begin{tabular}{rl}
			\includegraphics[width=5cm]{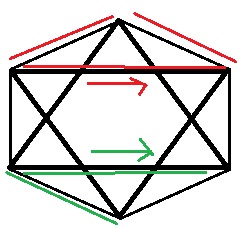}
		\end{tabular}
	\end{center}
	\caption{A non-symmetric hexagonal manifold is formed by joining two different trigonal manifolds.The direction of 
	photons on two trigonal manifolds are reverse to each other, however  they can't cancel the effect of each other. }
\end{figure*}
%%%%%%%%%%%%%%%%%%%%%%%%%%%%%%%%%%%%%%%%%%%%%%%%%%%%%%%%%%%%%%%%%%%%%%%%%%%%%%%%%%%%%%% 

%%%%%%%%%%%%%%%%%%%%%%%%%%%%%%%%%%%%%%%%%%%%%%%%%%% Figure 3  %%%%%%%%%%%%%%%%%%%%%%%%%%%%
\begin{figure*}[thbp]
	\begin{center}
		\begin{tabular}{rl}
			\includegraphics[width=5cm]{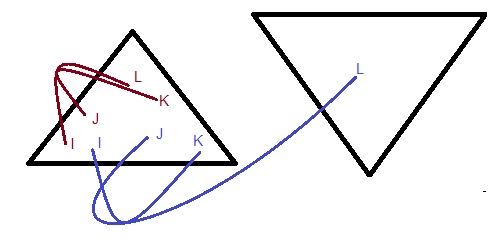}
		\end{tabular}
	\end{center}
	\caption{GG-fields and Cs-fields  are formed by joining exchanged photons. }
\end{figure*}
%%%%%%%%%%%%%%%%%%%%%%%%%%%%%%%%%%%%%%%%%%%%%%%%%%%%%%%%%%%%%%%%%%%%%%%%%%%%%%%%%%%%%%% 

%%%%%%%%%%%%%%%%%%%%%%%%%%%%%%%%%%%%%%%%%%%%%%%%%%% Figure 4  %%%%%%%%%%%%%%%%%%%%%%%%%%%%
\begin{figure*}[thbp]
	\begin{center}
		\begin{tabular}{rl}
			\includegraphics[width=5cm]{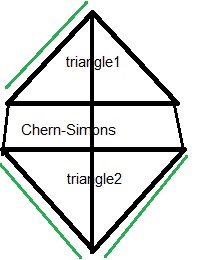}
		\end{tabular}
	\end{center}
	\caption{A hexagonal diode consisted of two trigonal manifolds are connected by a Chern-Simons manifold.}
\end{figure*}
%%%%%%%%%%%%%%%%%%%%%%%%%%%%%%%%%%%%%%%%%%%%%%%%%%%%%%%%%%%%%%%%%%%%%%%%%%%%%%%%%%%%%%% 

\section{The BIonic diode }\label{o2}
In this section, we will construct the BIonic diode by connecting a pentagonal and a heptagonal manifold by a Chern-Simons manifold. 
The  energy and Hamiltonian of pentagonal manifold has a reverse sign in respect to  the energy and the Hamiltonian of heptagonal 
manifold. Consequently, pentagonal manifold absorbs electrons and heptagonal molecules repels them. \\

A pentagonal manifold can be built of two trigonal manifolds with a common vertex (See Figure 5). Consequently, both of trigonal 
manifolds have a common photonic field. To avoid of calculating this photon for two times, we remove it from one of trigonal 
manifolds. We have:

\begin{eqnarray}
&& S_{5}^{tot}=S_{3}-\bar{S}_{2}\nonumber\\&& H_{5}^{tot}=H_{3}-\bar{H}_{2}
\label{EQ9}
\end{eqnarray}

Following the mechanism  in previous section, we obtain following actions:

\begin{eqnarray}
&& S_{3}=-T_{tri} \int d^{3}\sigma \sqrt{\eta^{ab}
	g_{MN}\partial_{a}X^{M}\partial_{b}X^{N}+2\pi
	l_{s}^{2}(\sum_{n=1}^{3}\frac{1}{n!}(-\frac{F_{1}..F_{n}}{\beta^{2}}))
	)}
\nonumber\\&& \bar{S}_{2}=-T_{tri} \int d^{3}\sigma \sqrt{\eta^{ab}
	g_{MN}\partial_{a}\bar{X}^{M}\partial_{b}\bar{X}^{N}+2\pi
	l_{s}^{2}(\sum_{n=1}^{2}\frac{1}{n!}(-\frac{\bar{F}_{1}..\bar{F}_{n}}{\beta^{2}}))
	)}
\label{EQ10}
\end{eqnarray} 

and following Hamiltonians:

\begin{eqnarray}
&&H_{3}=4\pi T_{tri}\int d\sigma_{3} d\sigma_{2} d\sigma_{1} \sqrt{1+\eta^{ab}g_{MN}\partial_{a}X^{M}\partial_{b}X^{N}}O^{3}_{tot}
\nonumber\\ &&
\neq \bar{H}_{2}=4\pi T_{tri}\int d\bar{\sigma}_{3} d\bar{\sigma}_{2} d\bar{\sigma}_{1} \sqrt{1+\eta^{ab}g_{MN}\partial_{a}\bar{X}^{M}
\partial_{b}\bar{X}^{N}}\bar{O}^{2}_{tot} \nonumber\\ && O^{3}_{tot}=\sqrt{1+\frac{k^{2}_{3}}{O_{2}\sigma^{4}_{3}}}
\sqrt{1+\frac{k^{2}_{2}}{O_{1}\sigma^{4}_{2}}}\sqrt{1+\frac{k^{2}_{1}}{\sigma^{4}_{1}}}\nonumber\\
&&\bar{O}^{2}_{tot}=\bar{O}_{1}\sqrt{1+\frac{k^{2}_{2}}{\bar{O}_{1}\bar{\sigma}^{4}_{2}}}
\label{EQ11}
\end{eqnarray}

After doing some algebra on the above Hamiltonians and using the mechanism in (\ref{EQ6}), we obtain:

\begin{eqnarray}
&& H_{tot}^{5} = H_{3} - \bar{H}_{2} \approx
\nonumber\\&& -(4\pi T_{tri}) [1+(\frac{2\pi
	l_{s}^{2}}{\beta^{2}}) [\bar{G}_{IJKL}\bar{G}^{IJKL}-G_{IJKL}G^{IJKL}]\nonumber\\&& +
(\frac{(2\pi 
	l_{s}^{2})^{2}}{\beta^{4}})(\psi_{1,\downarrow}\psi_{1,\uparrow}\psi_{2,\downarrow}\psi_2,{\uparrow})'[ Cs \bar{G}_{IJKL}
\bar{G}^{IJKL}-\bar{Cs} G_{IJKL}
G^{IJKL} ]\nonumber\\&&
\nonumber\\&& -(\frac{2\pi
	l_{s}^{2}}{\beta^{2}})^{3} [\bar{G}_{IJKLMN}\bar{G}^{IJKLMN}]\nonumber\\&& -
(\frac{(2\pi 
	l_{s}^{2})^{2}}{\beta^{4}})^{3}(\psi_{1,\downarrow}\psi_{1,\uparrow}\psi_{2,\downarrow}\psi_2,{\uparrow})'[ Cs \bar{G}_{IJKLMN}
\bar{G}^{IJKLMN} ]+.........]
\label{EQ12}
\end{eqnarray}

This equation shows that similar to the hexagonal manifold, the exchanged photons between trigonal manifolds in pentagonal manifold 
form Chern-Simons fields, however the pentagonal manifold has less Chern-Simons and GG-fields. This is because that the pentagonal 
manifold has less sides in respect to hexagonal manifold and consequently, it's exchanged photons are less.

%%%%%%%%%%%%%%%%%%%%%%%%%%%%%%%%%%%%%%%%%%%%%%%%%%% Figure 5  %%%%%%%%%%%%%%%%%%%%%%%%%%%%
\begin{figure*}[thbp]
	\begin{center}
		\begin{tabular}{rl}
			\includegraphics[width=5cm]{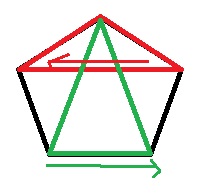}
		\end{tabular}
	\end{center}
	\caption{A pentagonal manifold is formed by joining two trigonal manifolds with a common vertex. }
\end{figure*}
%%%%%%%%%%%%%%%%%%%%%%%%%%%%%%%%%%%%%%%%%%%%%%%%%%%%%%%%%%%%%%%%%%%%%%%%%%%%%%%%%%%%%%% 

Also, using  the Hamiltonians in equation (\ref{EQ11})
and assuming all coordinates are the same (
$\sigma_{1} = \sigma_{2} =\sigma_{3}$), we obtain:

\begin{eqnarray}
&& E_{tot}^{5} = H_{3} - \bar{H}_{2} \approx
4k\pi T_{tri} [\frac{1}{\sigma^{5}}-
\frac{1}{\sigma^{3}}]
\nonumber\\&& F=-
\frac{\partial E}{\partial \sigma}=4k\pi T_{tri} [\frac{1}{\sigma^{6}}-
\frac{1}{\sigma^{4}}] \ll 0
\label{EQ13}
\end{eqnarray}

This equation shows that the force which is applied by a pentagonal manifold to an electron is attractive. Thus this manifold 
attracts the electrons.  In fact, a pentagonal manifold should be connected by another manifold and obtain the needed electrons. 
In next step, we want to consider the behaviour of heptagonal manifolds. \\

A pentagonal manifold is formed by joining three trigonal manifolds which have two comon vertexes. These two 
trigonal manifolds build a system with four vertexes and four fields (See figure 6 and figure 7). Thus, we can write:

\begin{eqnarray}
&& S_{7}^{tot}=S_{3}-\bar{S}_{3-3}\nonumber\\&& H_{7}^{tot}=H_{3}-\bar{H}_{3-3}
\label{EQ14}
\end{eqnarray}

Using the method in previous section, we obtain following actions

\begin{eqnarray}
&& S_{7}=-T_{tri} \int d^{3}\sigma \sqrt{\eta^{ab}
	g_{MN}\partial_{a}X^{M}\partial_{b}X^{N}+2\pi
	l_{s}^{2}(\sum_{n=1}^{3}\frac{1}{n!}(-\frac{F_{1}..F_{n}}{\beta^{2}}))
	)}
\nonumber\\&& \bar{S}_{3-3}=-T_{tri} \int d^{3}\sigma \sqrt{\eta^{ab}
	g_{MN}\partial_{a}\bar{X}^{M}\partial_{b}\bar{X}^{N}+2\pi
	l_{s}^{2}(\sum_{n=1}^{4}\frac{1}{n!}(-\frac{\bar{F}_{1}..\bar{F}_{n}}{\beta^{2}}))
	)}
\label{EQ15}
\end{eqnarray} 

and following Hamiltonians:

\begin{eqnarray}
&&H_{3}=4\pi T_{tri}\int d\sigma_{3} d\sigma_{2} d\sigma_{1} \sqrt{1+\eta^{ab}g_{MN}\partial_{a}X^{M}\partial_{b}X^{N}}O^{3}_{tot}
\nonumber\\ &&
\neq \bar{H}_{3-3}=4\pi T_{tri}\int d\bar{\sigma}_{3} d\bar{\sigma}_{2} d\bar{\sigma}_{1} \sqrt{1+\eta^{ab}g_{MN}\partial_{a}\bar{X}^{M}
\partial_{b}\bar{X}^{N}}\bar{O}^{3-3}_{tot} \nonumber\\ && 
O^{3}_{tot}=\sqrt{1+\frac{k^{2}_{3}}{O_{2}\sigma^{4}_{3}}}\sqrt{1+\frac{k^{2}_{2}}{O_{1}\sigma^{4}_{2}}}\sqrt{1+\frac{k^{2}_{1}}
{\sigma^{4}_{1}}}\nonumber\\
&&\bar{O}^{3-3}_{tot}=\sqrt{1+\frac{k^{2}_{4}}{\bar{O}_{3}\bar{\sigma}^{4}_{4}}}\sqrt{1+\frac{k^{2}_{3}}{\bar{O}_{2}\bar{\sigma}^{4}_{3}}}
\sqrt{1+\frac{k^{2}_{2}}{\bar{O}_{1}\bar{\sigma}^{4}_{2}}}\sqrt{1+\frac{k^{2}_{1}}{\bar{\sigma}^{4}_{1}}}
\label{EQ16}
\end{eqnarray}

Using the Taylor series in the above Hamiltonians and applying the  method in (\ref{EQ6}) yields:

\begin{eqnarray}
&& H_{tot}^{7} = H_{3} - \bar{H}_{3-3} \approx
\nonumber\\&& -(4\pi T_{tri}) [1+(\frac{2\pi
	l_{s}^{2}}{\beta^{2}}) [\bar{G}_{IJKL}\bar{G}^{IJKL}-G_{IJKL}G^{IJKL}]\nonumber\\&& +
(\frac{(2\pi 
	l_{s}^{2})^{2}}{\beta^{4}})(\psi_{1,\downarrow}\psi_{1,\uparrow}\psi_{2,\downarrow}\psi_2,{\uparrow})'[ Cs \bar{G}_{IJKL}
\bar{G}^{IJKL}-\bar{Cs} G_{IJKL}
G^{IJKL} ]\nonumber\\&& -(\frac{2\pi
	l_{s}^{2}}{\beta^{2}})^{3} [\bar{G}_{IJKLMN}\bar{G}^{IJKLMN}-G_{IJKLMN}G^{IJKLMN}]\nonumber\\&& -
(\frac{(2\pi 
	l_{s}^{2})^{2}}{\beta^{4}})^{3}(\psi_{1,\downarrow}\psi_{1,\uparrow}\psi_{2,\downarrow}\psi_2,{\uparrow})'[ Cs \bar{G}_{IJKLMN}
\bar{G}^{IJKLMN}-\bar{Cs} G_{IJKLMN}
G^{IJKLMN} ]
\nonumber\\&& -(\frac{2\pi
	l_{s}^{2}}{\beta^{2}})^{4} [G_{IJKLMNYZ}G^{IJKLMNYZ}]\nonumber\\&& -
(\frac{(2\pi 
	l_{s}^{2})^{2}}{\beta^{4}})^{4}(\psi_{1,\downarrow}\psi_{1,\uparrow}\psi_{2,\downarrow}\psi_2,{\uparrow})'[
\bar{Cs} G_{IJKLMNYZ}
G^{IJKLMNYZ} ]+.........]
\label{EQ17}
\end{eqnarray}

This equation indicates that like the hexagonal and pentagonal manifold, the exchanged photons between trigonal manifolds
in heptagonal manifold form Chern-Simons fields, however the heptagonal manifold has more Chern-Simons and GG-fields. This 
is because that the heptagonal manifold has more sides in respect to hexagonal manifold and consequently, it's exchanged 
photons are more.

Similar to pentagonal manifold, using  the Hamiltonians in equation (\ref{EQ16})
and assuming all coordinates are the same ($\sigma_{1} = \sigma_{2} =\sigma_{3}$), we obtain:

\begin{eqnarray}
&& E_{tot}^{7} = H_{3} - \bar{H}_{3-3} \approx
4k\pi T_{tri} [\frac{1}{\sigma^{5}}-
\frac{1}{\sigma^{9}}]
\nonumber\\&& F=-
\frac{\partial E}{\partial \sigma}=4k\pi T_{tri} [\frac{1}{\sigma^{6}}-
\frac{1}{\sigma^{10}}] \gg 0
\label{EQ18}
\end{eqnarray}

This equation indicates that the force which is applied by a heptagonal manifold to an electron is repulsive. Thus this 
manifold repel the electrons.  In fact, a heptagonal manifold should be connected by a pentagonal manifold and gives 
the extra electrons to it.  
%%%%%%%%%%%%%%%%%%%%%%%%%%%%%%%%%%%%%%%%%%%%%%%%%%% Figure 7  %%%%%%%%%%%%%%%%%%%%%%%%%%%%
\begin{figure*}[thbp]
	\begin{center}
		\begin{tabular}{rl}
			\includegraphics[width=5cm]{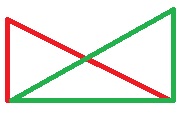}
		\end{tabular}
	\end{center}
	\caption{ Two trigonal manifolds with two comon vertexes. }
\end{figure*}
%%%%%%%%%%%%%%%%%%%%%%%%%%%%%%%%%%%%%%%%%%%%%%%%%%%%%%%%%%%%%%%%%%%%%%%%%%%%%%%%%%%%%%% 

%%%%%%%%%%%%%%%%%%%%%%%%%%%%%%%%%%%%%%%%%%%%%%%%%%% Figure 6  %%%%%%%%%%%%%%%%%%%%%%%%%%%%
\begin{figure*}[thbp]
	\begin{center}
		\begin{tabular}{rl}
			\includegraphics[width=5cm]{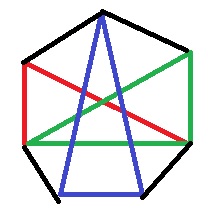}
		\end{tabular}
	\end{center}
	\caption{ A heptagonal manifold is formed by joining three trigonal manifolds which two of them have two comon vertexes. }
\end{figure*}
%%%%%%%%%%%%%%%%%%%%%%%%%%%%%%%%%%%%%%%%%%%%%%%%%%%%%%%%%%%%%%%%%%%%%%%%%%%%%%%%%%%%%%% 
 A BIonic diode can be constructed from a pentagonal manifold which is connected to heptagonal manifold via a
  Chern-Simons fields (See figure 8). Using the Hamiltonians in (\ref{EQ12} and \ref{EQ17}), we obtain:
  
  \begin{eqnarray}
  && H_{tot}^{Diode} =  H_{tot}^{5} +  H_{tot}^{7} \approx
  \nonumber\\&& -(4\pi T_{tri}) [ -(\frac{2\pi
  	l_{s}^{2}}{\beta^{2}})^{3} [\bar{G}_{IJKLMN}\bar{G}^{IJKLMN}]\nonumber\\&& -
  (\frac{(2\pi 
  	l_{s}^{2})^{2}}{\beta^{4}})^{3}(\psi_{1,\downarrow}\psi_{1,\uparrow}\psi_{2,\downarrow}\psi_2,{\uparrow})'[ Cs \bar{G}_{IJKLMN}
  \bar{G}^{IJKLMN} ]
  \nonumber\\&&-(\frac{2\pi
  	l_{s}^{2}}{\beta^{2}})^{4} [G_{IJKLMNYZ}G^{IJKLMNYZ}]\nonumber\\&& -
  (\frac{(2\pi 
  	l_{s}^{2})^{2}}{\beta^{4}})^{4}(\psi_{1,\downarrow}\psi_{1,\uparrow}\psi_{2,\downarrow}\psi_2,{\uparrow})'[
  \bar{Cs} G_{IJKLMNYZ}
  G^{IJKLMNYZ} ]+.........]
  \label{EQ19}
  \end{eqnarray}
  
  This equation shows that the Hamiltonian of the BIonic diode includes terms with 6 and 8 indices. This means that the rank of 
  Cs-GG terms in a pentagonal-heptagonal diode is more than the rank of Cs-GG terms in a hexagonal diode. In fact, in 
  pentagonal-heptagonal diode more photonic fields are exchanged and stability of system is more than the hexagonal diode.  \\   
  
  Using equations (\ref{EQ8},\ref{EQ11}, \ref{EQ16}), we can obtain the photonic mass in a BIonic diode:
  
   \begin{eqnarray}
  &&[m_{photon}^{2}]=\frac{1}{l_{Chern-Simons}} \int_{\sigma}^{\infty}
  d\sigma'(\frac{[O_{Diode}(\sigma)^{2}-
  	\bar{O}_{Diode}(\sigma)^{2}]}{[O_{Diode}(\sigma_{0})^{2}-\bar{O}_{Diode}(\sigma_{0})^{2}]}-1)^{-1/2} \label{EQ20}
  \end{eqnarray}

where

\begin{eqnarray}
&& O_{Diode}=2O^{3}_{tot}
-O^{3-3}_{tot}-O^{2}_{tot}
\label{EQ21}
\end{eqnarray}

In a pentagonal-heptagonal BIonic diode, the photonic mass depends not only on the separation distance between manifolds but also 
on the shape and topology of trigonal manifolds which construct manifolds. It is clear that for a small gap between manifolds, 
coupling of photons to electrons on the Chern-Simons manifold increases and they become massive.

%%%%%%%%%%%%%%%%%%%%%%%%%%%%%%%%%%%%%%%%%%%%%%%%%%% Figure 8  %%%%%%%%%%%%%%%%%%%%%%%%%%%%
\begin{figure*}[thbp]
	\begin{center}
		\begin{tabular}{rl}
			\includegraphics[width=5cm]{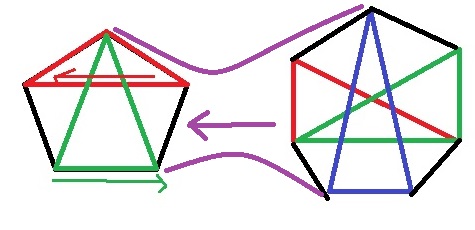}
		\end{tabular}
	\end{center}
	\caption{ A BIonic diode can be constructed from a pentagonal manifold which is connected to heptagonal manifold via a 
	Chern-Simons manifold. }
\end{figure*}
%%%%%%%%%%%%%%%%%%%%%%%%%%%%%%%%%%%%%%%%%%%%%%%%%%%%%%%%%%%%%%%%%%%%%%%%%%%%%%%%%%%%%%% 

\section{Summary} \label{sum}
In this paper, we have considered the formation and the evolutions of BIonic diodes on the polygonal manifolds. For example, we 
have shown that a hexagonal BIonic diode can be constructed by two non-similar trigonal manifolds. Photons which  are exchanged 
between trigonal manifolds, form Chern-Simons fields which live on a Chern-Simons manifold. The hexagonal BIons interact 
with each other via connecting two Chern-Simons manifolds. For a hexagonal manifold with similar trigonal manifolds, exchanged 
photons cancel the effect of each other and the energy and also the length of Chern-Simons manifold becomes zero. These manifolds 
are stable and cannot interact with each other. If  the symmetry of hexagonal manifolds is broken, another polygonal manifolds like 
heptagonal and pentagonal manifolds are formed. Phtonos that are exchanged between these manifolds form two Chern-Simons fields which 
live on two Chern-Simons manifolds. These manifolds connect to each other and construct a BIonic diode. Photons that move via this 
manifold, lead to the motion of electrons from heptagonal side to pentagonal side. These photons are massive and their mass depends 
of the angles between atoms and length of gap between two manifolds.
%%%%%%%%%%%%%%%%%%%%%%%%%%%%%%%%%%%%%%%%%%%%%%%%%%%%%%%%%%%%%%%%%%%%%
\section*{Acknowledgements}
\noindent The work of Alireza Sepehri has been supported financially by the Research Institute for Astronomy and Astrophysics
of Maragha (RIAAM),Iran under the Research Project NO.1/5237-77. A. Pradhan also acknowledges IUCAA, Pune, India for providing 
facility during a visit under a Visiting Associateship where a part of this paper has been done.
%%%%%%%%%%%%%%%%%%%%%%%%%%%%%%%%%%%%%%%%%%%%%%%%%%%%%%%%%%%%%%%%%%%%%%%%%%%%%%%%%%%%


\begin{thebibliography}{1}
	
\bibitem{q2}
A.D. Bartolomeo, Physics Reports 606 (2016) 1-58.
\bibitem{q3}
T.J. Echtermeyer, P.S. Nene, M. Trushin, R.V. Gorbachev, A.L. Eiden, S. Milana, Z. Sun, J. Schliemann, E. Lidorikis, 
K.S. Novoselov, A. C. Ferrari, Nano Lett. 14 (2014) 3733.
\bibitem{q4}
T. Mueller, F. Xia, P. Avouris,  Nature Photon. 4 (2010) 297-301.
\bibitem{q5}
C.E.P. Villegas, P.B. Mendonça, A. Rocha, Scientific Reports 4 (2014) 6579.
\bibitem{q6}
R.R. Nair, P. Blake, J.R. Blake, R. Zan, S. Anissimova, U. Bangert, A.P. Golovanov, S.V. Morozov, T. Latychevskaia, 
A.K. Geim, K.S. Novoselov, Appl. Phys. Lett. 97 (2010) 153102. 
\bibitem{q7}
S. Parui, R. Ruiter, P.J. Zomer, M. Wojtaszek, B.J. van Wees, T. Banerjee, Journal of Applied Physics 116 (2014) 244505.
\bibitem{q8}Alireza Sepehri, Umesh Kumar Sharma, Anirudh Pradhan, In press in Physics Letters A, (2017)\\ Alireza Sepehri, Int. J. Geom. Methods Mod. Phys. https://doi.org/10.1142/S0219887817501523,  arXiv:1707.04283.\\A. Sepehri, R. Pincak, A.F. Ali, Eur. Phys. J. B 89 (2016) 250.\\ 
A. Sepehri, R. Pincak, K. Bamba, S. Capozziello, E. N. Saridakis, 
International Journal of Modern Physics D, 1750094 (2017),  arXiv:1607.01499 [gr-qc].
\bibitem{qq8} Marcelo Vieira, Sergei Sergeenkov, Claudio Furtado, Phys. Rev. A 96, 013852 (2017).

\bibitem{D3} 
G. Grignani, T. Harmark, A. Marini, N.A. Obers, M. Orselli, JHEP 1106 (2011) 058.
\bibitem{Df}Alireza Sepehri, Farook Rahaman, Salvatore Capozziello, Ahmed Farag Ali, Anirudh Pradhan, Eur.Phys.J. C76 (2016) no.5, 231.
\bibitem{b1}
P. Horava, E. Witten, Nucl. Phys. B 460 (1996) 506--524, arXiv:hep-th/9510209; P. Horava, E. Witten, 
Nucl. Phys. B 475 (1996) 94-114, arXiv:hep-th/9603142.
	
\end{thebibliography}
\end{document}